\documentstyle[PASJadd,psbox]{PASJ95}
\draft

\newcommand{\vol}{}
\newcommand{\no}{}
\newcommand{\lett}{}
\newcommand{\spage}{}
\newcommand{\rdate}{}
\newcommand{\adate}{}

\begin{document}

\title{Metal Rich Plasma at the Center Portion \\ of the Cygnus Loop}

\author{Emi {\sc Miyata}\thanks{CREST, Japan Science and
	Technology Corporation (JST)}, \  Hiroshi {\sc Tsunemi}$^*$,
	Takayoshi {\sc Kohmura}, and Seiji {\sc Suzuki} \\
	{\it Department of Earth and Space Science,
	Graduate School of Science, Osaka University }\\
	{\it 1-1, Machikaneyama, Toyonaka, Osaka, 560} \\
	{\it E-mail(EM) : miyata@ess.sci.osaka-u.ac.jp} \\
	and \\
	Shiomi {\sc Kumagai} \\
	{\it Department of Physics, College of Science and Technology, 
	Nihon University} \\
	{\it Kanda-Surugadai 1-8, Chiyoda-ku, Tokyo 101}\\}

\abst{We observed the center portion of the Cygnus Loop supernova remnant
with the ASCA observatory. The X-ray spectrum of the center portion
was significantly different from that obtained at the North-East (NE)
limb. The emission lines from Si and S were quite strong while those
of O and the continuum emission were similar to those obtained at the
NE limb.  Based on the spectral analysis, Si and S emission lines
originated from a high-kTe and low ionization plasma whereas O and
most of the continuum emission arose from a low-kTe and high
ionization plasma.  We suppose that Si and S emitting gas are present
at the interior of the Loop while O lines and continuum emission
mainly arise from the shell region. Therefore, we subtracted the
spectrum of the NE limb from that of the center.  Obtained abundances
of Si, S, and Fe were 4 $\pm$ 1, 6 $\pm$ 2, and ${1.3}^{+0.6}_{-0.3}$
times higher than those of the cosmic abundances, respectively,
and are $\sim$40 times richer than those obtained at the NE limb.
These facts strongly support that some of the crude ejecta must be
left at the center portion of the Cygnus Loop.  The low abundance of
Fe relative to Si and S suggests a type II SN with a massive
progenitor star as the origin of the Cygnus Loop.  }

\kword{Supernovae, supernova remnants --- Abundances --- X-rays: spectra}

\maketitle
\thispagestyle{headings}

\section{Introduction}

The heavy elements present in the hot gas in galaxies are thought to
be supplied both by stellar winds and by supernovae (SNe). Theoretical
calculations of the nucleosynthesis of the SNe have been improved
greatly in the past ten years (e.g., Thielemann, Nomoto, \& Hashimoto
1990; Thielemann, Nomoto, \& Hashimoto 1996, hereafter TNH).  In
contrast, the observational evidences of metal rich gas have been
obtained for only a limited number of young supernova remnants (SNRs).

Ku et al. (1984) first performed spatially-resolved spectroscopy on
the Cygnus Loop with the Einstein observatory. They constructed an
X-ray image with spatial resolution of $\sim$1$^\prime$ in the energy
region of 0.1--4.0 keV. The X-ray image clearly showed
limb-brightening structure.  They noticed that kTe at the limb was
generally lower than that at the center.  Charles, Kahn, \& McKee
(1985) found that kTe gradually increased toward the center from the
vicinity of the limb.  Tsunemi et al. (1988) observed the whole Cygnus
Loop with the Gas Scintillation Proportional Counters on board the
Tenma satellite, which possessed roughly twice better energy
resolution than that of the proportional counter (Koyama et al. 1984). 
They detected emission lines from highly ionized Si and S. They fitted
the X-ray spectrum in combination with the spectrum obtained with a
previous rocket observation (Inoue et al. 1979, 1980). They found that
a two-component NEI model could reproduce the X-ray spectra.  Together
with Charles et al. (1985), Tsunemi et al. (1988) suggested that
tenuous high-kTe plasma was sitting in the interior region which was
surrounded by dense low-kTe plasma. Such high-kTe component was
confirmed by Hatsukade \& Tsunemi (1990) with the Ginga satellite.

Due to its high sensitivity, high energy resolution, and wide energy
band, the ASCA observatory has opened a new window for the physics of
hot plasmas like SNRs. We can perform the plasma diagnostics in detail
by using the X-ray CCD cameras.  So far, the abundances of heavy
elements have been well determined for many young SNRs with the ASCA
observatory (see review by Tsunemi \& Miyata 1997).  For evolved SNRs,
it is difficult to determine the abundances due to low-kTe, low
surface brightness, and large interstellar absorption. Furthermore,
the swept up interstellar medium (ISM) dominates the ejecta mass from
the progenitor star. This makes it difficult to detect the ejecta.

The Cygnus Loop is an evolved SNR and one of the best studied
SNRs. Since the Cygnus Loop is roughly 8 degrees away from the
Galactic plane, the neutral hydrogen column (${\rm N_H}$) is only a
few$\times {10}^{20}\ {\rm cm}^{-2}$. This enables us to detect
emission lines from O and to determine kTe with high accuracy. The
mass of the swept up ISM is estimated to be roughly 100 ${\rm
M_{\odot}}$ since the average density of the ISM is 0.2 ${\rm
cm}^{-3}$ and the mean radius is 18.8 pc (Ku et al.  1984). In
comparison, the mass of the ejecta ranges from one to several tens
${\rm M_{\odot}}$. Thus, the ejecta are submerged under the sea of
the ISM.  However, due to its high surface brightness and large
apparent size, we can perform spatially-resolved analysis for the
Cygnus Loop. This enables us to search ejecta in detail if ejecta have
not yet mixed well with the ISM.

The first observation of the Cygnus Loop with the ASCA observatory was
performed by Miyata et al. (1994; hereafter MTPK) on the NE limb. They
found that kTe increased toward the center whereas
log\hspace{0.2mm}($\tau$) decreased. They determined the abundances of
heavy elements from O to Fe and found that metals were deficient at
the NE limb.

In this {\it paper}, we present the observational results of the
center portion of the Cygnus Loop observed with the ASCA observatory.
The X-ray spectrum obtained from the center portion was quite
different from that obtained at the NE limb.  We studied the plasma
diagnostics of the interior of the Cygnus Loop and nucleosynthesis at
the SN, apart from dense shell regions.

\section{Observations and Results}

We observed the center portion of the Cygnus Loop with the ASCA
observatory for $\sim$15~ks on Apr. 20--21, 1993 during the ASCA
performance-verification phase. The ASCA observatory, the fourth
Japanese X-ray satellite (Tanaka, Inoue, \& Holt 1994), is equipped
with four X-ray telescopes (XRT; Serlemitsos et al. 1995) that
simultaneously feed four focal plane instruments.  The telescopes have
a point spread function (PSF) with a half power diameter of about
3$^\prime$.  The focal plane detectors are two Solid-state Imaging
Spectrometers (SIS; Burke et al. 1991) and two Gas Imaging
Spectrometers (GIS; Ohashi et al. 1996; Makishima et al. 1996). The
SISs have a FOV of $22^\prime \times 22 ^\prime$ and an energy
resolution of approximately 60 eV full width at half maximum below
1~keV.  The GISs have a higher effective area above 3 keV and larger
FOV than the SISs. We focused on the SIS data for the spectra analysis
and used the GIS data to investigate the spatial variation.

Figure~1 shows the location of our SIS FOV superimposed on the X-ray
surface brightness map of the Loop (Aschenbach 1994) with a black
square.  It is centered at $\alpha = {\rm 20^h 51^m 56^s}$ and $\delta
= 31^{\rm d} 09^\prime 02^{\prime\prime}$ (2000).  The location of the
observation by MTPK toward the NE limb is also shown in this figure.

There were several correction factors (Echo, DFE, CTI, and RDD) in the
SIS data analysis (Dotani et al. 1995; Dotani et al. 1997).  As
mentioned in Dotani et al.  (1997), the energy resolution of the SIS
gets worse due to the RDD effects, particularly for 4 CCD mode
data. Our data were obtained just after the launch and the energy
degradation caused by the RDD effects could be negligibly small. Thus,
we corrected only Echo, DFE and CTI.  Since these effects could be
corrected only for the Faint mode data, we focused only on the Faint
mode data for the spectrum analysis.  The observing time of the Faint
mode data was $\sim$5.6 ks after screening the data.  We subtracted
blank sky spectra (NEP and Lynx field regions) as the background
since we estimated the contribution of the Galactic X-ray background
to be negligibly small (Koyama et al. 1986). The count rate in our FOV
was $\sim$8.8 c ${\rm s^{-1}\ SIS^{-1}}$.

Another serious background component for an extended source is stray
light (Serlemitsos et al. 1995). We estimated the contamination of the
center portion of the Cygnus Loop caused by the bright shell regions.
The count rate of the stray light from the shell regions was roughly
0.2 c ${\rm s^{-1}}$, which was only 2~\% of the total count rate of
the center portion. Therefore, we can safely ignore the contamination
from the bright shell regions. See the appendix for more detail.

\subsection{Spectrum Fitting}

The X-ray spectrum of our FOV is shown in figure~2, which was the sum
of the data from two SISs.  K emission lines of O, Ne, and Mg were
relatively weak while those of Si and S, and L emission lines of Fe
were quite strong compared with those obtained at the NE limb
(MTPK). We detected three emission lines for Si: Si$\;${\small XIII}
K$\alpha$, K$\beta$, and Si$\;${\small XIV} K$\alpha$.  Using the line
intensity ratios of these emission lines, we can determine the
ionization condition of the X-ray emitting plasma as mentioned in
section~\ref{sec:combination}

\subsubsection{NEI model (Fit I)}

We first applied the NEI model coded by K. Masai (Masai 1984) to our
data (hereafter we refer this fitting as Fit I). The NEI model is
characterized both with kTe and with $\tau\equiv {\rm n_e t}$, where
${\rm n_e}$ is an electron density [${\rm cm}^{-3}$] and t is an
elapsed time [s] after the shock heating.  In this model, we
calculated the evolution of the ionization with a constant kTe.  kTe
mainly depends on the shape of the continuum emission while $\tau$ can
be determined either by line intensity ratios of a given element or by
the line center energies of He-like ions.  The abundances of O, Ne,
Mg, Si, S, Fe, and Ni were set to be free parameters.  That of He was
set to be a cosmic value (Allen 1973) whereas those of C and N were
fixed to that of O. ${\rm N_H}$ was fixed to 4${\rm\times 10^{20}\ 
cm^{-2}}$ (Inoue et al. 1979; Kahn et al. 1980). The best fit curve is
shown in figure~2 and derived parameters are summarized in table~1.
This model was far from an acceptable fit, with a \mbox{\rm reduced
$\chi^2$} of 18 (degrees of freedom (dof) = 77). From the statistical
point of view, a large discrepancy was found around 0.6 keV, where
emission lines from O$\;${\small VII} and O$\;${\small VIII} were
present.  The best fit model could not reproduce the line intensity
ratio between O$\;${\small VII} and O$\;${\small VIII}. The data
requested higher intensity from O$\;${\small VII} and lower intensity
from O$\;${\small VIII} than the model provided. This suggested lower
ionization state and/or lower kTe than that expected from the best fit
model. The other large discrepancy was found around Fe--L blends in
the energy range of 0.8--1.2 keV, where Fe--L emission lines must be
dominant. We should note that the best fit model also could not
reproduce the line intensity ratios between Si$\;${\small XIII}
K$\alpha$, K$\beta$, and Si$\;${\small XIV} K$\alpha$. The data
preferred higher intensity from Si$\;${\small XIII} K$\beta$ than that
of the best fit model, which suggested higher kTe.

Through the model fitting by using the NEI model, we found that no
model with a single component could reproduce the line intensity
ratios of O and Si lines simultaneously. This means that a
multi-component plasma is present along the line of sight in our
FOV. This is natural since Charles et al. (1985) and Sauvageot,
Decourchelle, \& Tsunemi (1995) found gradients in kTe and MTPK found
gradients both in kTe and in log\hspace{0.2mm}($\tau$) along the
radius in the Cygnus Loop.  From the theoretical point of view, we can
expect such gradients for SNRs in the adiabatic phase (e.g., Itoh 1979)
even if we consider the Coulomb heating of electron gas (Shklovskii
1962; Ito 1978). These observational results and theoretical
calculations suggest that the value of kTe increases toward the
explosion center. Thus, we suppose that emission lines of Si mainly
arise from the inner region while those of O arise from the outer
region.

\subsubsection{Combination model (Fit II)}\label{sec:combination}

	Next, we applied a model of thermal bremsstrahlung and
Gaussian line profiles, a combination model as noted in MTPK, in order
to evaluate the emission lines (hereafter we refer this fitting as Fit
II).  Free parameters were center energy, line width, normalization of
each Gaussian line, kTe and normalization of thermal
bremsstrahlung. Applying the F-test with a significance level of 99 \%
to determine the number of Gaussian lines and continuum emission, we
found that 13 Gaussian lines and two components of continuum emission
with different kTe significantly improved $\chi^2$ value.  The
significance level of the last component added (the 2.24~keV line) was
$\ge$99~\%. No additional component could improve the $\chi^2$ value
even at the 90~\% confidence level.  The best fit curve is shown in
figure~3. The best fit parameters for Gaussian line profiles are
summarized in table~2. We identified each line and the results are
shown in this table. The line width of each Gaussian line profile
except Fe--L blends, Si$\;${\small XIII} K$\alpha$, and S$\;${\small
XV} K$\alpha$ is negligibly small, taking into account the energy
resolution of the SIS. Obtained kTe values are 0.24 $\pm$ 0.01 and
${1.1}^{+3}_{-0.5}$ keV and the reduced $\chi^2$ is 1.0 with dof of
57.  The equivalent width of each emission line is also given in this
table.

We found that the continuum emission consisted of two components with
different kTe. Such high-kTe component was first detected by the Tenma
satellite (Tsunemi et al. 1988) and confirmed by the Ginga satellite
(Hatsukade \& Tsunemi 1990).  Based on their observation of the whole
Cygnus Loop, these authors suggested that the tenuous high-kTe plasma
fills the interior region which was surrounded by the dense low-kTe
plasma. Due to the projection effect, we can easily imagine that we
detected at least two types of plasma with different kTe and different
density when we observed at the center portion.

The obtained kTe for the low-kTe component is well within the range
obtained at the NE limb (0.23--0.30 keV). The emission measures (${\rm
n_e}^{2}$L [pc ${\rm cm}^{-6}$]: L is the plasma length along the line
of sight [pc]) of the two components were 14 $\pm$ 1 and
(${5}^{+16}_{-4})\times {10}^{-2}$ pc ${\rm cm}^{-6}$ for the low-kTe
component and the high-kTe component, respectively. Therefore, we
suppose that most of the continuum emission arises from low-kTe and
high density shell region.

We calculated the line intensity ratios of Si$\;${\small XIII}
K$\alpha$ to Si$\;${\small XIII} K$\beta$ and Si$\;${\small XIV}
K$\alpha$ to Si$\;${\small XIII} K$\alpha$. The former is sensitive to
kTe and the latter is sensitive both to kTe and to
log\hspace{0.2mm}($\tau$). Thus, we can determine the plasma condition
for Si.  Figure~4 shows the confidence contours of these intensity
ratios on a log\hspace{0.2mm}($\tau$)--kTe plane, based on the
calculations performed by Masai (1984). The area shaded with
right-down lines shows the 90~\% confidence area of Si$\;${\small
XIII} K$\alpha$ / Si$\;${\small XIII} K$\beta$ while that with
right-up lines shows the 90~\% confidence area of Si$\;${\small XIV}
K$\alpha$ / Si$\;${\small XIII} K$\alpha$. We can restrict the plasma
condition of Si as kTe $\ge$ 1.5~keV and log\hspace{0.2mm}($\tau$)
$\le$ 10.6, suggesting significant departure from the collisional
ionization equilibrium (CIE) condition. This high kTe value is
consistent with that of the continuum emission of the high kTe
component.  These results suggest that there is a hot unevolved plasma
somewhere along the line of sight.

For S emission lines, we could not detect S$\;${\small XV} K$\beta$ or
higher ionization ions. Therefore, we could determine neither kTe nor
log\hspace{0.2mm}($\tau$) by using the line intensity ratios. As shown
by Tsunemi et al. (1986), the center energies of the lines from the
ions of heavy elements up to He-like depend both on kTe and on
log\hspace{0.2mm}($\tau$). The variation of the line center energy for
S is large compared with other elements (O--Si) under the plasma
conditions we are concerned with. We calculated the apparent line
center energy for S in a log\hspace{0.2mm}($\tau$)--kTe plane as shown
in figure~5.  We included lines in the energy range of 2.3 to 2.5~keV
from highly ionized ions of S up to He-like ions.  The 90~\% error
region for the apparent line center energy of S is shown by the shaded
area. From this figure, we found that the kTe of the S emitting plasma
was quite high ($\ge$ 1.5 keV) and similar to that of Si.  The
ionization timescale can be determined to be $\ge$ 10.6.  When we
considered the systematic errors of 10 eV in the determination of the
energy scale, kTe and log\hspace{0.2mm}($\tau$) for S emitting plasma
could be restricted to $\ge$ 1 keV and $\ge$ 10.5,
respectively. Therefore, kTe $\simeq$ 1.5 keV and
log\hspace{0.2mm}($\tau$) $\simeq$ 10.6 is acceptable both for Si and
for S.

We next studied the equivalent widths as shown in table~2.  We found
remarkable large equivalent widths for Si and S lines for which we
investigated whether the plasma with the cosmic abundance could
account or not.  The equivalent width of the emission line depends on
kTe, log\hspace{0.2mm}($\tau$), and the elemental abundance.
Figures~6--7 show the equivalent widths of Si and S in a
log\hspace{0.2mm}($\tau$)--kTe plane, assuming the cosmic
abundance. We summed all emission lines from Si in the energy range
from 1.839 to 1.866 keV to calculate the figure 6. For figure 7, we
considered all emission lines from S in the energy range from 2.3 to
2.5 keV. The maximum values of the equivalent width in kTe $\le$ 100
keV are roughly 0.9~keV and 0.5~keV for Si and S, respectively. The
kTe range in figure 6 surely covers the kTe range we have to consider. 
Therefore, the above values can be considered as maximum for our
purpose. These maximum values are much smaller than those we obtained.
This indicates that the abundances of Si and S are higher than those
of cosmic values.

We estimated the most plausible values and lower limits for the
abundances of Si and S. As shown in figures 6--7, the equivalent
widths of Si and S are $\sim$ 0.6 keV and $\sim$ 0.3 keV for the
cosmic plasma with kTe=1.5 keV, log\hspace{0.2mm}($\tau$)=10.6.
Therefore, we calculated the most plausible abundances by dividing the
observed equivalent widths by these values and got 4 and 8 times
larger than the cosmic abundances for Si and S, respectively. In the
same way, we calculated the lower limits of abundances by dividing
observed equivalent widths by the maximum values in the case of cosmic
abundances. Obtained lower limits for Si and S were 3 and 5.
Therefore, we established the over-abundance for Si and S at the
center portion.  Adding to the studies of line intensity ratios, we
suppose that hot unevolved plasma containing rich Si and S is sitting
at the center portion of the Cygnus Loop along the line of sight.

We also calculated the line intensity ratio for O emission lines.  The
line intensity ratio of O$\;${\small VIII} to O$\;${\small VII} was
0.47 $\pm$ 0.04. This value was consistent with the average value
obtained by MTPK (0.15--0.80).  This means that the plasma condition
of O at the center portion is similar to that at the NE limb. Next, we
compared the line intensities at the center portion with those at the
NE limb.  Due to the projection effect, if we observe at the center
portion, we detect emission both from the shell region and from the
interior region.  At the shell region, we could expect low-kTe
continuum and strong O emission lines as we detected at the NE
limb. Since the plasma length at the center was different from that at
the NE limb, we compared intensities of O lines by using the
equivalent width. At the NE limb, the equivalent widths for
O$\;${\small VIII} and O$\;${\small VII} were 0.17 and 0.20 keV,
respectively. In table~2, we calculated equivalent width of each
emission line by using both continuum emission components. If we
calculated equivalent widths by using the low-kTe component solely, we
obtained 0.18 and 0.20 keV for O$\;${\small VIII} and O$\;${\small
VII}, respectively. These values are in good agreement with those
obtained at the NE limb.  This fact strongly suggests that all of O
emission lines were accounted for by the low-kTe component, which would
be present at the shell region.  Therefore, we can safely assume that
the emission lines of O mainly come from the shell region which is
seen at the center portion due to the projection effect.

In summary of Fit II, we obtained high kTe values both for Si and for
S based on the studies of ratios of emission line intensities. We
confirmed that Si and S emitting gas was over-abundant based on the
studies of equivalent widths. On the other hand, most of the continuum
emission arose from the low-kTe component and the obtained kTe value
was similar to that obtained at the NE limb (MTPK). The line intensity
ratio of O$\;${\small VIII} to O$\;${\small VII} was also consistent
with that obtained at the NE limb.  These results suggest that Si and
S arise from the interior plasma with high-kTe ($\simeq$ 1.5 keV)
showing significant departure from the CIE condition
(log\hspace{0.2mm}($\tau$) $\simeq$ 10.6) while most of the continuum
emission and O arise from the low-kTe plasma in the shell region.

\subsection{Comparison with the NE Limb}\label{sec:comparison}

Figure~8 shows the X-ray spectra obtained both at the center portion
and at the NE limb. We normalized these spectra by equalizing line
intensities of O.  The emission lines from Si, S, and Fe in the center
portion are much stronger than those in the NE limb, whereas the line
intensity ratios of O$\;${\small VIII} to O$\;${\small VII} are
similar, as indicated in section 2.1.2. To evaluate the emission from
the interior region (hereafter we refer to this region as the `core'
region), we subtracted the spectrum of the NE limb from the center
portion by equalizing the intensity of O lines.  Figure~9 shows the
limb-subtracted spectrum.

Si, S, and Fe--L blends were prominent in the core spectrum. We again
applied the NEI model to this spectrum (hereafter we refer this
fitting as Fit III). The abundance of He was fixed to the cosmic
value. Since those of C, N, and O could not be determined, we fixed
them to 0.25 which was obtained in the Fit I.  Free parameters were
kTe, log\hspace{0.2mm}($\tau$), emission measure, abundances of Ne,
Mg, Si, S, Fe, and Ni.  The best fit curve is shown in figure~9 and
the best fit parameters are summarized in table~1. Whereas fits were
significantly improved, this model was not yet acceptable from the
statistical point of view (reduced $\chi^2$ of 2.8 with dof of
71). Large discrepancy could be found again at the energy region of
Fe--L ($\sim$1.2 keV).  Liedahl, Osterheld, \& Goldstein (1995)
recalculated emissivities of Fe--L lines and found deviations from the
previously known model, especially for the atomic code of
Fe$\;${\small XXIII}--Fe$\;${\small XXIV} (3$\rightarrow$2). The NEI
model we currently used should still be tuned up to take account of
the new atomic code. However, except for the 1.2 keV energy region,
other Fe--L lines ($\le$ 1 keV) were well reproduced with the current
model, and the abundance of Fe must be determined with these lines
from the statistical point of view.  Therefore, we expect that
abundance of Fe would be reliable.

The obtained kTe was consistent with that of the continuum emission
for the high-kTe component in Fit II but was slightly lower than that
obtained from the studies of line intensity ratios of Si. We should
note that we obtained higher kTe (1.1 $\pm$ 0.3) if we fitted the core
spectrum with the same model in the energy range of 1.5--4~keV. In
this case, the obtained abundances were consistent with those obtained
with Fit III because the statistical errors became large.

Obtained abundances of Si and S were much larger than those of the
cosmic abundances.  The abundance of Fe was consistent with the cosmic
value.  These values were roughly 40 times larger than those obtained
at the NE limb (MTPK). These results indicate that hot and low
ionization condition plasma with rich Si, S, and Fe is present
along the line of sight at the center portion of the Cygnus Loop.

\section{Discussion}

\subsection{Concentration of Heavy Elements at the Core Region}

We applied the NEI model to the core spectrum and obtained an
over-abundance for Si, S, and Fe. Compared to the abundances obtained
at the NE limb (MTPK), those obtained at the core region were a factor
of $\sim$40 larger. The plasma condition of the Si and S emitting
plasma is in high-kTe and in low ionization state, suggesting its
presence at the interior region of the Cygnus Loop.  These facts
strongly support that the X-ray emitting plasma at the core region is
the ejecta in origin.  Chevalier (1974) and Mansfield \& Salpeter
(1974) suggested that some amount of mass of a progenitor star would
remain at the core region of the remnant even in an evolved SNR like
the Cygnus Loop. Our results also indicate that the mixing of ejecta
into the ISM did not occur effectively.

We then investigated the extent of the Si, S, and Fe emitting region.
Fe emission lines are in the energy region of 0.7--1.3~keV where we
expect many emission lines from other elements. On the contrary, in
the energy region of 1.8--2.5~keV, we expect only Si and S emission
lines. Since MTPK found the deficient abundance of Fe at the NE limb,
we suppose that emission lines from Fe would arise from the similar
region to those of Si and S.  We extracted the GIS image with the
narrow energy range of 1.8--2.5~keV where GIS possesses large
effective area. Emission in this energy band arose from a circular
region with a radius of $\sim$9$^\prime$ $\approx$ 2~pc as shown in
figure~10. Assuming a spherical uniformly emitting region (filling
factor was set to unity), we calculated the density and mass of heavy
elements listed in table~3.

\subsection{Comparison with Theoretical Calculations}

We compared the derived abundances of Si, S, and Fe with those
predicted by the theoretical calculations of nucleosynthesis in the
SNe.  After the explosive processing of heavy elements by the passing
of the supernova shock, heavy elements are formed in a so-called
``onion-skin'' structure.  Light elements like C, N, or O are sitting
in outer layers whereas heavier elements like Si, S, or Fe are
produced at the inner region (TNH). It is not reasonable to compare
our results with the mean abundance of nucleosynthesis for an entire
star (table 3 in TNH), since we observed only the core region of the
Cygnus Loop.  Therefore, we integrated the abundances of the heavy
elements from an explosion center (Mr = ${\rm M_0}$; Mr is mass radius
and normalized by the stellar mass) toward some mass radius (Mr =
${\rm M_1}$), considering the mass fraction diagrams as shown in
figure 7 in Nomoto, Thielemann \& Yokoi (1984) for type Ia (W7 model)
and figure 1 in TNH for type II with various progenitor star
masses. For the type Ia model, ${\rm M_0}$ is zero while for the type
II model, ${\rm M_0}$ is fixed to the mass cut radius described in
table 4 in TNH.  The neutron star or black hole would form inside the
mass cut.  We summed up $^{56}$Ni, $^{57}$Ni, and Fe to calculate the
Fe mass.  Figure~11 shows the comparison of the derived mass ratios of
S/Si and Fe/Si (shown by dotted lines) with the model calculations of
the type Ia model (a) and type II models with progenitor masses of
15${\rm M_{\odot}}$(b), 20${\rm M_{\odot}}$(c), and 25${\rm
M_{\odot}}$(d).  In general, if we integrate from ${\rm M_0}$ to some
mass radius, we can trace along the solid lines from upper right to
lower left in each diagram.

The type Ia model predicts the mass of Fe to be much larger than that of
Si and could not reproduce our results even if we set ${\rm M_1}$ to
be 1.4${\rm M_{\odot}}$.

For the type II model, TNH cannot precisely determine the location of
the mass cut, with the result that the Fe abundance is rather
uncertain.  The Fe mass predicted by the type II model with 20${\rm
M_{\odot}}$ was well determined based on the observation of
SN1987A. Since the radioactive decays of $^{56}$Ni$\rightarrow
^{56}$Co$\rightarrow ^{56}$Fe power the light curve, the amount of
$^{56}$Ni was determined to be 0.075${\rm M_{\odot}}$ (e.g., Arnett
et al. 1989). In the same way, the Fe mass in the type II model with
15${\rm M_{\odot}}$ was also confirmed since the derived Fe mass
based on the light curve of 1993J was 0.1${\rm M_{\odot}}$ (e.g.,
Nomoto et al. 1993).

The type II model with progenitor mass less than 20${\rm M_{\odot}}$
can reproduce the mass ratio of S to Si we obtained. However, the
total mass of Fe is much larger than that of Si even if we integrate
from the mass cut radius to the outermost layer. Therefore, the
observed mass ratio of Fe to Si cannot be explained with these models.

For the type II model with 25${\rm M_{\odot}}$, we can explain the
obtained mass ratio of S to Si as well as Fe to Si with a reasonable
value of ${\rm M_1}$ (2.0--3.9${\rm M_{\odot}}$).  When we integrate
from ${\rm M_0 = 1.77M_{\odot}}$ to ${\rm M_1 = 2.0 M_{\odot}}$, the
predicted masses of Si, S, and Fe are 0.063, 0.035, and 0.041${\rm
M_{\odot}}$.  Therefore, we can explain the derived masses of Si, S,
and Fe with the 25${\rm M_{\odot}}$ model if roughly 1~\% of the
ejecta sitting at 1.77${\rm M_{\odot}} \le$Mr$\le2.0{\rm M_{\odot}}$
is still present at the core region of the Cygnus Loop.  We also
estimated the acceptable range of ${\rm M_0}$. We found $1.77 \le {\rm
M_0} \le 1.85$ can reproduce the mass ratios we obtained.  This result
suggests that the mass inside 1.77--$1.85{\rm M_{\odot}}$ will
collapse into a neutron star.

Our results support the model proposed by McCray \& Snow (1979). McKee
\& Cowie (1975) suggested that there would exist dense clouds in the
vicinity of the Cygnus Loop in order to explain the differences in the
shock velocities seen at optical wavelengths (Minkowski 1958)
and in the soft X-ray region (Gorenstein et al. 1971). McCray \& Snow
(1979) proposed that such dense clouds would be associated with the
formation of a stellar wind bubble by an early-type progenitor star,
meaning type II SN origin of the Cygnus Loop.  Charles et al. (1985)
also suggested type II origin based on the inhomogeneities in the
X-ray image. It should be noted that we were the first to determine
the stellar type of the progenitor of the Cygnus Loop supernova based
on the elemental abundance ratios of the ejecta as determined from
X-ray lines.

The most ejecta have left our FOV.  Considering the metal deficiency
at the NE limb, we suppose that most ejecta, including all light
elements like O or Mg as well as Si, S, and Fe, are still inside
the shell and isolated from the ISM. If it is the case, the contact
discontinuity, which separates the ejecta from the shocked ISM, may be
seen as the abundance discontinuity.

A SN explosion with a progenitor mass of $\ge 10 {\rm M_{\odot}}$ is
thought to form a neutron star. We expect that a neutron star with a
mass of 1.77--${\rm M_{\odot}}$ should have born at the SN
explosion that created the Cygnus Loop.  So far, there is no
observational evidence of a neutron star associated with the Cygnus
Loop. Further deep observations in the radio and X-ray wavelengths are
strongly encouraged to confirm the massive star origin of the Cygnus
Loop.

\section{Conclusion}

We observed the center portion of the Cygnus Loop supernova remnant
with the X-ray CCD cameras on board the ASCA observatory.  We
confirmed a significant departure from a collisional ionization
equilibrium condition at the center portion. Obtained abundances at
the core region were larger than those of cosmic abundances for Si, S.
Moreover, abundances of Si, S, and Fe were $\sim$40 times larger than
those obtained at the NE limb. This strongly supports the hypothesis
that the X-ray emitting plasma in the core is ejecta in origin.
Although previous X-ray observations could not detect any signature of
ejecta, we could find it by using both the imaging capability and the
high energy resolving power of the ASCA observatory.

Obtained abundances can be compared with theoretical calculations of
nucleosynthesis by a SN explosion. Since we only observed the core
region, we integrated the model calculations from the explosion center
to some mass radius. For the explosion models for type Ia SN or for
type II with the progenitor mass less than 20 ${\rm M_{\odot}}$, the
abundance of Fe is much larger than that of Si or S even if we
integrate to the outermost radius.  On the other hand, the abundances
of Si, S, and Fe can be explained with the type II model with 25${\rm
M_{\odot}}$ if we integrate these abundances from the mass cut radius
1.77${\rm M_{\odot}}$ to $2 \sim 3.9 {\rm M_{\odot}}$.  Therefore, our
results suggest the massive progenitor origin of the Cygnus Loop and
the existence of the stellar remnant associated with the Cygnus
Loop. \par

\vspace{1pc}\par
We would like to acknowledge Prof. Nomoto and Drs. T. Suzuki and
K. Iwamoto for fruitful discussions about the theoretical aspects of
the SNRs at conference of Thermonuclear Supernovae.  Dr. B. Aschenbach
kindly gave us the whole X-ray image of the Cygnus Loop obtained with
the ROSAT all-sky survey.  We thank Dr. Thielemann to use his data on
nucleosynthesis of SNe.  We would like to thank the anonymous referee
for her or his detailed comments and suggestions which greatly
improved this paper.

We are greatful to all the members of ASCA team for their
contributions to the fabrication of the apparatus, the operation of
ASCA, and the data acquisition. We thank the members of {\tt
ASCA\_ANL/Sim ASCA} software team.

\section*{Appendix. \ Effects of stray light}

We estimated the effects of stray light.  For the ASCA X-ray
telescope, XRT, some X-rays coming from an off axis angle of
0.5--1.5$^\circ$ can reach the focal plane (Serlemitsos et
al. 1995). There are mainly four kinds of stray paths expected: 1)
single reflection by primary mirrors (referred as Primary); 2) single
reflection by secondary mirrors (Secondary); 3) direct
(No-reflection); and 4) a few times reflections by front and back
surfaces of mirrors (Abnormal) (Tsusaka et al. 1994). There are no
bright X-ray stars around the Cygnus Loop.  However, if we observe the
center portion of the Loop, the contamination by emission from the
bright shell region might be important.

We simulated stray light effects in the frame work of {\tt
ASCA\_ANL/Sim ASCA} ver. 0.9. We used the code of the ray tracing,
{\tt XRTraytr} ver 2.1 developed by the XRT team (and modified by us). 
This code simulated the average performance of the four XRTs and did
not consider the characteristics of each telescope.

We assumed limb-brightening features with the inner and outer radii of
72$^\prime$ and 84$^\prime$, which were similar to the global
structure of the Cygnus Loop as shown in figure~1. We also estimated
the stray light as well as the usual double reflections (Normal) when
we observed the center portion of the Cygnus Loop.  Under this
configuration, we confirmed no X-rays from the bright shell region of
the Cygnus Loop reached the focal plane as Normal.

Figure~12 shows the results of our simulations for the SIS.  The
coordinate system is {\tt DETX/Y} which is a detector coordinate
system.  The horizontal and vertical gaps are due to the physical
spaces between the CCD chips of the SIS. We rebinned each image by
4$\times$4 pixels.  We generated 7 million 1~keV photons for each
stray component.  The fraction of each component was written below
each figure. In total, 0.1~\% of the incident photons came into the
whole FOV of the SIS as stray light shown in the right-lower. We
should note that the relative intensity of the stray light increases
toward the edge of the detector.

We did not yet obtain the whole image of the Cygnus Loop with the ASCA
observatory. The whole image of the Cygnus Loop obtained with the
ROSAT all-sky survey showed the inhomogeneities of the shell regions
(Aschenbach 1994). We, here, estimated the contamination for the worst
case. We assumed the count rate of the shell region per each
22$^\prime$ square to be the same as that of the NE limb and 13 ${\rm
s^{-1}}$ which was one of the brightest regions in the ROSAT energy
band and also the brightest region in the ASCA energy band ever
observed. Thus, the total count rate of the shell region could be
estimated as roughly 170 ${\rm s^{-1}}$, resulting in the contamination
to the center portion of 0.17 ${\rm s^{-1}}$. This count rate was
similar to that of the background including both the internal
background and the CXB (Gendreau et al. 1995). This value was only
2~\% of the total count of the center portion and we concluded that we
could safely ignore the contamination from the bright shell regions.

\clearpage

\section*{References}
\re Allen, C.W. 1973, Astrophysical Quantities 3rd ed.
	(The Athlone Press), p.30
\re Arnett, W.D., Bahcall, J.N., Kirshner, R.P., Woosley, S.E.
	1989, ARA\&A, 27, 629
\re Aschenbach, B. 1994, New Horizon of X-ray Astronomy,
	eds. F. Makino \& T. Ohashi (Universal Academy Press inc.), p.103
\re Burke, B.E., Mountain, R.W., Daniels, P.J., Cooper, M.J.,
	\& Dolat, V.S. 1991, IEEE Trans., ED-38, 1069
\re Charles, P.A., Kahn, S. M., \& McKee, C. F. 1985, ApJ, 295, 456
\re Chevalier, R.A. 1974, ApJ, 188, 501
\re Cox, D.P. 1972, ApJ, 178, 169
\re Dotani, T., Yamashita, A., Rasmussen, A., \& the SIS team,
	1995, ASCA News Letter 3, p.25
\re Dotani, T., Yamashita, A., Ezuka, H., Takahashi, K., Crew, G.,
	Mukai, K., \& the SIS team, 1997, ASCA News Letter 5, p.14
\re Gendreau, K.C. Mushotzky, R., Fabian, A.C., Holt, S.S.,
	Kii, T., Serlemitsos, P.J., Ogasaka, Y., Tanaka, Y.
	et al. 1995, PASJ, 47, L5
\re Gorenstein, P., Harris, B., Gursky, H., \& Giacconi, R.
	1971, Science, 172, 369
\re Hatsukade, I., \& Tsunemi, H. 1990, ApJ, 362, 566
\re Inoue, H., Koyama, K., Matsuoka, M., Ohashi, T., Tanaka,
	Y., \& Tsunemi, H. 1979, X-Ray Astronomy,
	eds. W.A. Baity \& L.E. Peterson (Pergamon Press, Oxford), p.309
\re Inoue, H., Koyama, K., Matsuoka, M., Ohashi, T., Tanaka, T.,
	\& Tsunemi, H. 1980, ApJ, 238, 886
\re Itoh, H. 1978, PASJ, 30, 489; Erattum, 31, 429
\re Itoh, H. 1979, PASJ, 31, 541
\re Kahn, S.M., Charles, P.A., Bowyer, S., \& Blissett, R.J.
	1980, ApJ, 242, L19
\re Koyama, K., Ikegami, T., Inoue, H., Kawai, N., Makishima, K.,
	Matsuoka, M. Mitsuda, K., Murakami, T. et al. 1984, PASJ, 36, 659
\re Koyama, K., Makishima, K., Tanaka, Y., \& Tsunemi, H.
	1986, PASJ, 38, 121
\re Ku, W.H.-M., Kahn, S.M., Pisarski, R., \& Long, K.S. 1984, 
	ApJ, 278, 615
\re Liedahl, D.A., Osterheld, A.L.,
	\& Goldstein, W.H. 1995, ApJ, 438, L115
\re Makishima, K., Tashiro, M., Ebisawa, K., Ezawa, H.,
	Fukazawa, Y., Gunji, S., Hirayama, M., Idesawa, E.
	1996, PASJ, 48, 171
\re Mansfield, V.N., \& Salpeter, E.E. 1974, ApJ, 190, 305
\re Masai, K. 1984, Ap\&SS, 98, 367
\re McCray, R., \& Snow, T.P. 1979, ARA\&A, 17, 213
\re McKee, C.F., \& Cowie, L.L. 1975, ApJ, 195, 715
\re Minkowski, R. 1958, Rev.Mod.Phys. 30, 1048
\re Miyata, E., Tsunemi, H., Pisarski, R., \& Kissel, S. E.
	1994, PASJ, 46, L101 (MTPK)
\re Nomoto, K., Thielemann, F.-K., \& Yokoi, K.
	1984, ApJ, 286, 644
\re Ohashi, T., Ebisawa, K., Fukazawa, Y., Hiyoshi, K., Horii, M.,
	Ikebe, Y., Ikeda, H., Inoue, H. et al. 1996, PASJ, 48, 157
\re Sauvageot, J.L., Decourchelle, A., \& Tsunemi, H. 1995,
	UV and X-ray Spectroscopy of Astrophysical and 
	Laboratory Plasmas, eds. K. Yamashita \& T. Watanabe
	(Universal Academy Press inc.), p.303
\re Serlemitsos, P.J., Jalota, L., Soong, Y., Kunieda, H.,
	Tawara, Y., Tsusaka, Y., Suzuki, H., Sakima, Y.
	et al. 1995, PASJ, 47, 105
\re Shklovskii, I.S. 1962, Sov.Astron., 6, 162
\re Tanaka, Y., Inoue, H., \& Holt, S.S. 1994, PASJ, 46, L37
\re Thielemann, F.-K., Nomoto, K., \& Hashimoto, M. 1990,
	Supernovae, Les Houches, Session LIV,
	eds. S.Bludman, R.Mochkovitch \& J.Zinn-Justin, p.629
\re Thielemann, F.-K., Nomoto, K., \& Hashimoto, M. 1996,
	ApJ, 460, 408 (TNH)
\re Tsunemi, H., Yamashita, K., Masai, K., Hayakawa, S., \&
	Koyama, K. 1986, ApJ, 306, 248
\re Tsunemi, H., Manabe, M., Yamashita, K., \& Koyama, K.
	1988, PASJ, 40, 449
\re Tsunemi, H., \& Miyata, E. 1997, Thermonuclear Supernovae,
	eds. P.Ruiz-Lapuente, R. Canal, \& J. Isern
	(Kluwer Academic Publishers), p.561
\re Tsusaka, Y., Suzuki, H., Awaki, H., Yamashita, K., Kunieda, H.,
	Tawara, Y., Ogasaka, Y., Uchibori, Y. et al. 1994, SPIE, 2011, 517

\newpage

\begin{fv}{1}{5cm}{The X-ray surface brightness map of the Cygnus Loop
	obtained with the ROSAT all-sky survey (Aschenbach 1994).
	The black squares show the FOV of the center portion
	and the NE limb (MTPK) as observed with the ASCA observatory.}
\end{fv}
\begin{fv}{2}{5cm}{X-ray spectrum of the center portion of the Cygnus Loop.
	Upper pannel shows the data points with $\pm$ 1 $\sigma$
	errors and the best fit curve with the NEI model (Fit I).
	Lower pannel shows the residuals of the fit in the unit of $\sigma$.}
\end{fv}
\begin{fv}{3}{5cm}{Same as figure 2, but for the combination model
	(Fit II).}
\end{fv}
\begin{fv}{4}{5cm}{The line intensity ratios of three Si lines
	in a log\hspace{0.2mm}($\tau$)--kTe plane.  The area shaded
        with right-down lines shows the 90 \% confidence area of
        Si$\;${\small XIII} K$\alpha$ / K$\beta$ while that with
        right-up lines shows the 90 \% confidence area of
        Si$\;${\small XIV} K$\alpha$ / Si$\;${\small XIII} K$\alpha$.}
\end{fv}
\begin{fv}{5}{5cm}{Center energy of the apparent S plotted
	in a log($\tau$)--kTe plane.  We considered the highly ionized
	ions up to He-like ions.  The shaded area
	shows the 90 \% confidence area for the line center energy of S.}
\end{fv}
\begin{fv}{6}{5cm}{The contour map of the equivalent widths
	[keV] of Si in a log\hspace{0.2mm}($\tau$)--kTe plane}
\end{fv}
\begin{fv}{7}{5cm}{Same as figure 6, but for S.}
\end{fv}
\begin{fv}{8}{5cm}{Spectrum of the center portion.
	For comparison, the spectrum obtained at the NE limb is also shown
	by equalizing at O line intensities.}
\end{fv}
\begin{fv}{9}{5cm}{The spectrum of the core region of the Cygnus Loop.
	The solid line shows the best fit curve of the NEI model
	(Fit III).}
\end{fv}
\begin{fv}{10}{5cm}{Narrow-band (1.8--2.5 keV) image
	obtained with the GIS. The FOV of the GIS is shown by
	solid circle.}
\end{fv}
\begin{fv}{11}{5cm}{Comparison of the derived mass ratios
	(S/Si and Fe/Si) with model calculations for nucleosynthesis
	of type Ia (a) and type II with 15${\rm M_{\odot}}$(b),
	20${\rm M_{\odot}}$(c), and
	25${\rm M_{\odot}}$(d) progenitor mass.  In the model calculations,
	we integrated abundances of Si, S, and Fe from ${\rm M_0}$ to
	${\rm M_1}$ and calculated S/Si and Fe/Si for each ${\rm
	M_1}$. Several ${\rm M_1}$ values are shown in diagrams in
	${\rm M_{\odot}}$ unit.
	Dotted lines show the regions we obtained in 90
	\% confidence level. }
\end{fv}
\begin{fv}{12}{5cm}{Simulated images of the stray light for No-reflect,
	Primary, Secondary, and Abnormal. The right-lower image is the
	sum for all components.  The vertical and horizontal gaps are
	due to the physical spaces between the CCD chips.}
\end{fv}

\newpage

\begin{table*}[t]
  \begin{center}
    Table~1. \hspace{4pt}Fitting results of the Masai model applied
	to the X-ray spectra extracted from the center portion
	and the core region  \\
  \end{center}
  \vspace{6pt}
  \begin{tabular*}{\columnwidth}{@{\hspace{\tabcolsep}
  \extracolsep{\fill}}p{15pc}ll}
  \hline\hline\\[-6pt]
	Parameter & Whole spectrum & Core spectrum \\
	& Fit I & Fit III  \\
	[4pt]\hline \\[-6pt]
	kTe [keV]\dotfill &  0.51 $\pm$ 0.01 & ${0.82}^{+0.09}_{-0.07}$\\
	log\hspace{0.2mm}($\tau$)\dotfill & 
		10.54 $\pm$ 0.03 & ${10.35}^{+0.06}_{-0.07}$\\
	Emission Measure [pc ${\rm cm}^{-6}$]${\rm ^a}$
		& 0.71 $\pm$ 0.06 & 0.10 $\pm$ 0.03 \\
	 \multicolumn{3}{l}{Abundance [relative to cosmic values]} \\
	\hspace{.65cm} C,N,O\dotfill & 0.25  $\pm$ 0.02 & 0.25 (fixed) \\
	\hspace{1cm} Ne\dotfill	& 0.46 $\pm$ 0.04 & ${0.2}^{+0.2}_{-0.1}$ \\
	\hspace{1cm} Mg\dotfill	& ${0.34}^{+0.05}_{-0.04}$ &
		${0.3}^{+0.3}_{-0.2}$ \\
	\hspace{1cm} Si\dotfill	& ${1.9}^{+0.2}_{-0.1}$
		& 4 $\pm$ 1 \\
	\hspace{1cm} S\dotfill & 5.4 $\pm$ 0.6 & 6 $\pm$ 2 \\
	\hspace{1cm} Fe\dotfill	& 0.46 $\pm$ 0.04 & ${1.3}^{+0.6}_{-0.3}$ \\
	\hspace{1cm} Ni\dotfill	& 7 $\pm$ 1 & ${5}^{+3}_{-2}$ \\
	reduced $\chi^2$ (dof) & 18 (77) &  2.8 (71) \\
  \hline
  \end{tabular*}
  \vspace{6pt}\par\noindent
  {\sc Noted}
	-- Quoted errors are at 90\% confidence level.\\
  ${\rm ^a}$ Emission measure in unit of pc$^3$ ${\rm cm}^{-6}$ was directly
  obtained from the spectrum analysis. We converted it to emission
  measure in unit of pc ${\rm cm}^{-6}$ to divide the obtained
  value by 22$^\prime\times$22$^\prime$ ($\simeq$ 24pc$^2$).
\end{table*}

\begin{table*}[t]
  \begin{center}
    Table~2. \hspace{4pt} The Observation Line Features (Fit II)\\
  \end{center}
  \vspace{6pt}
  \begin{tabular*}{\columnwidth}{@{\hspace{\tabcolsep}
  \extracolsep{\fill}}cccc}
  \hline\hline\\[-6pt]
	Observed energy & Flux & Ion species identified & Equivalent width \\
	$[$keV$]$ & $[10^{-3}$ photons ${\rm s^{-1}cm^{-2}}]$ & & $[$keV$]$ \\
	[4pt]\hline \\[-6pt]
	0.573 $\pm$ 0.002 & 40 $\pm$ 2 & O$\;${\small VII} 
         & 0.19 \\
	0.659 $\pm$ 0.003 & 19 $\pm$ 1 & O$\;${\small VIII}
         & 0.14 \\
	0.726 $\pm$ 0.003 & 14 $\pm$ 1 & Fe$\;${\small XVII}
         & 0.09 \\
	0.822 $\pm$ 0.003 & 35 $\pm$ 1 & Fe$\;${\small XVII}
         & 0.69 \\
	0.928 $\pm$ 0.003 & 6.0 $\pm$ 0.4 &
	 Ne$\;${\small IX}, Fe$\;${\small XVIII},
		Fe$\;${\small XIX}
         & 0.11 \\
	1.018 $\pm$ 0.004 & 2.9 $\pm$ 0.2 &
	 Ne$\;${\small IX}, Ne$\;${\small X},
	 Fe$\;${\small XVII}, Ni$\;${\small XX}
         & 0.12 \\
	1.121 $\pm$ 0.007 & 3.1 $\pm$ 0.2 & Fe$\;${\small XVII}
         & 0.29 \\
	1.251 $\pm$ 0.008 & 0.57 $\pm$ 0.09 &
	 Fe$\;${\small XIX}, Ni$\;${\small XIX}
         & 0.09 \\
	1.353 $\pm$ 0.005 & 0.54 $\pm$ 0.06& Mg$\;${\small XI}
         & 0.15 \\
	1.872 $\pm$ 0.002 & 1.02 $\pm$ 0.05 & Si$\;${\small XIII}
         & 2.3 \\
	2.07 $\pm$ 0.03 & 0.04 $\pm$ 0.02 & Si$\;${\small XIV}
         & 0.16 \\
	2.24 $\pm$ 0.01 & 0.12 $\pm$ 0.02 & Si$\;${\small XIII}
         & 0.81 \\
	2.463 $\pm$ 0.008 & 0.21 $\pm$ 0.03 & S$\;${\small XV}
                & 2.3 \\
  \hline
  \end{tabular*}
  \vspace{6pt}\par\noindent
  {\sc Noted}
	-- Systematic errors (about 10eV) are not included. \\
	-- Quoted errors are at 90\% confidence level.
\end{table*}

\begin{table*}[t]
  \begin{center}
    Table~3. \hspace{4pt} Densities and masses of heavy elements
	at the core region \\
  \end{center}
  \vspace{6pt}
  \begin{tabular*}{\columnwidth}{@{\hspace{\tabcolsep}
  \extracolsep{\fill}}p{10pc}cc}
  \hline\hline\\[-6pt]
	Elements & Density $[{10}^{-5} {\rm cm}^{-3}]$
	& Mass $[{10}^{-4} {\rm M_{\odot}}]$ \\
	[4pt]\hline \\[-6pt]
	Si \dotfill & 4 $\pm$ 1 & ${5}^{+2}_{-1}$ \\
	S \dotfill & ${2.7}^{+1.1}_{-0.8}$  & 4 $\pm$ 1 \\
	Fe \dotfill & ${1.4}^{+0.6}_{-0.4}$ & ${2.9}^{+1.4}_{-0.8}$ \\
  \hline
  \end{tabular*}
  \vspace{6pt}\par\noindent
  {\sc Noted}
	-- Quoted errors are at 90\% confidence level.
\end{table*}

\newpage

\begin{figure}[hptb]
  \centering
    \psbox{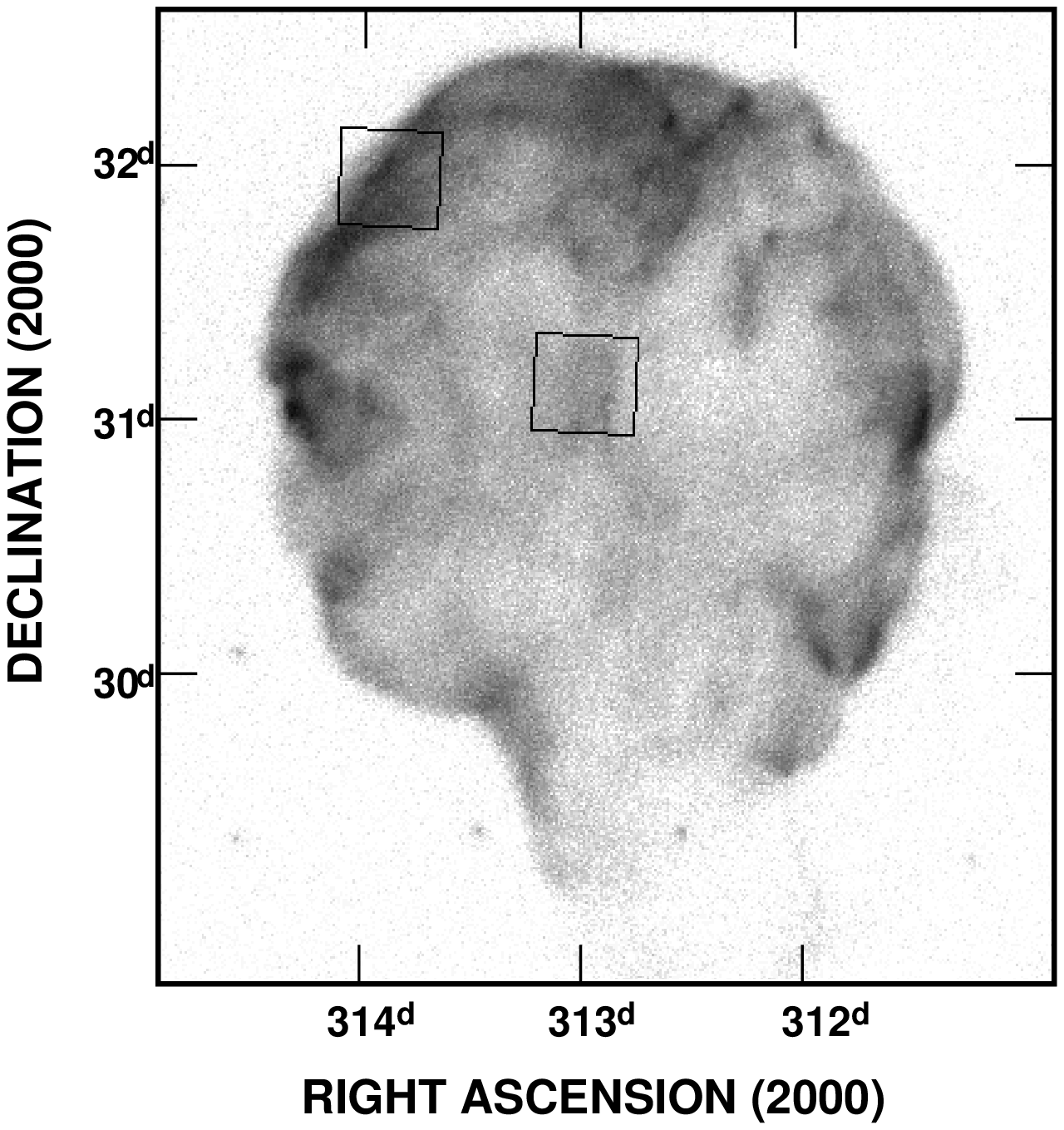}
\vspace{2cm}
\caption{}
\end{figure}

\begin{figure}[hptb]
  \centering
    \psbox[xsize=0.9#1]{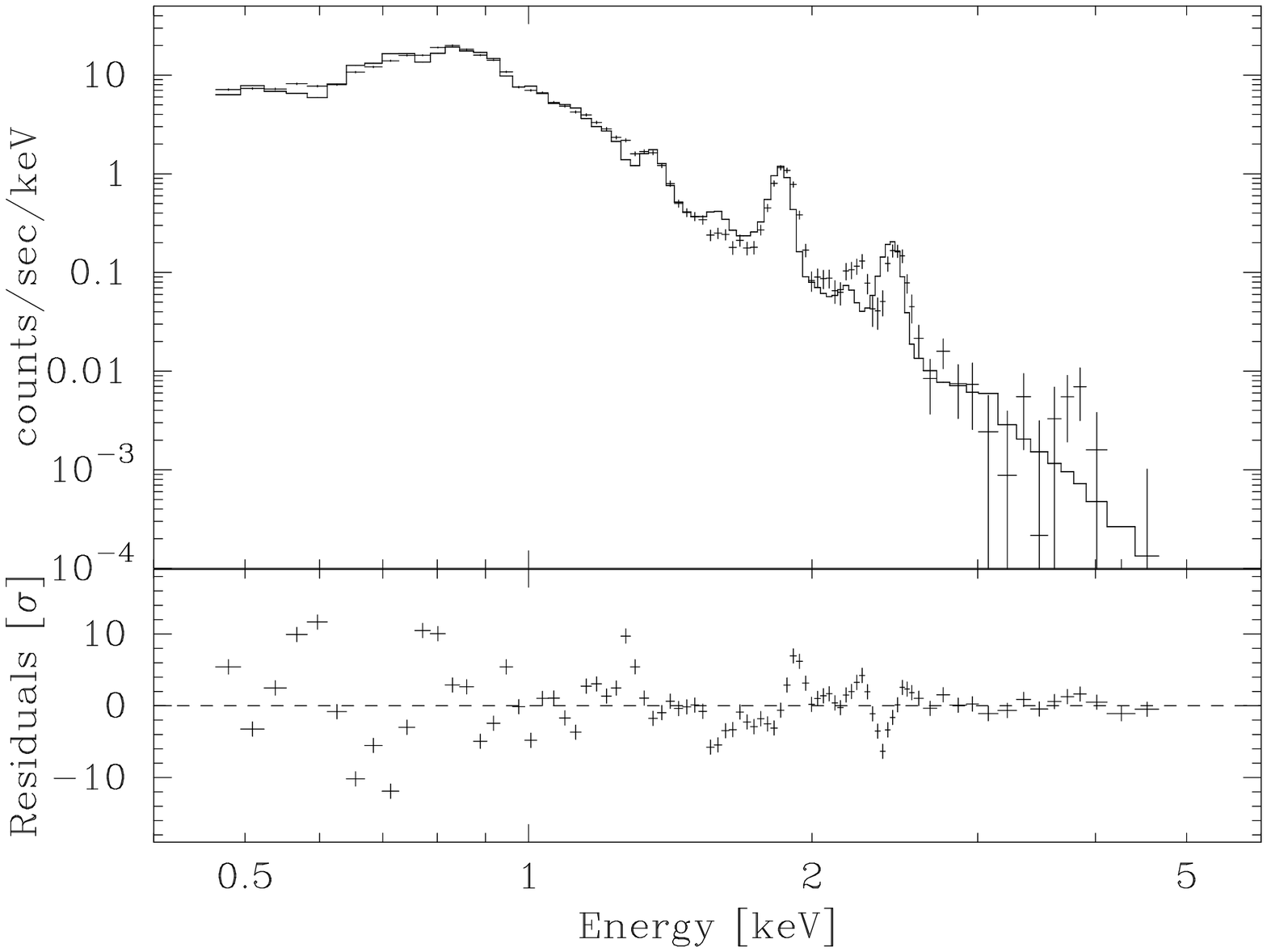}
\vspace{2cm}
\caption{}
\end{figure}

\begin{figure}[hptb]
  \centering
    \psbox[xsize=0.9#1]{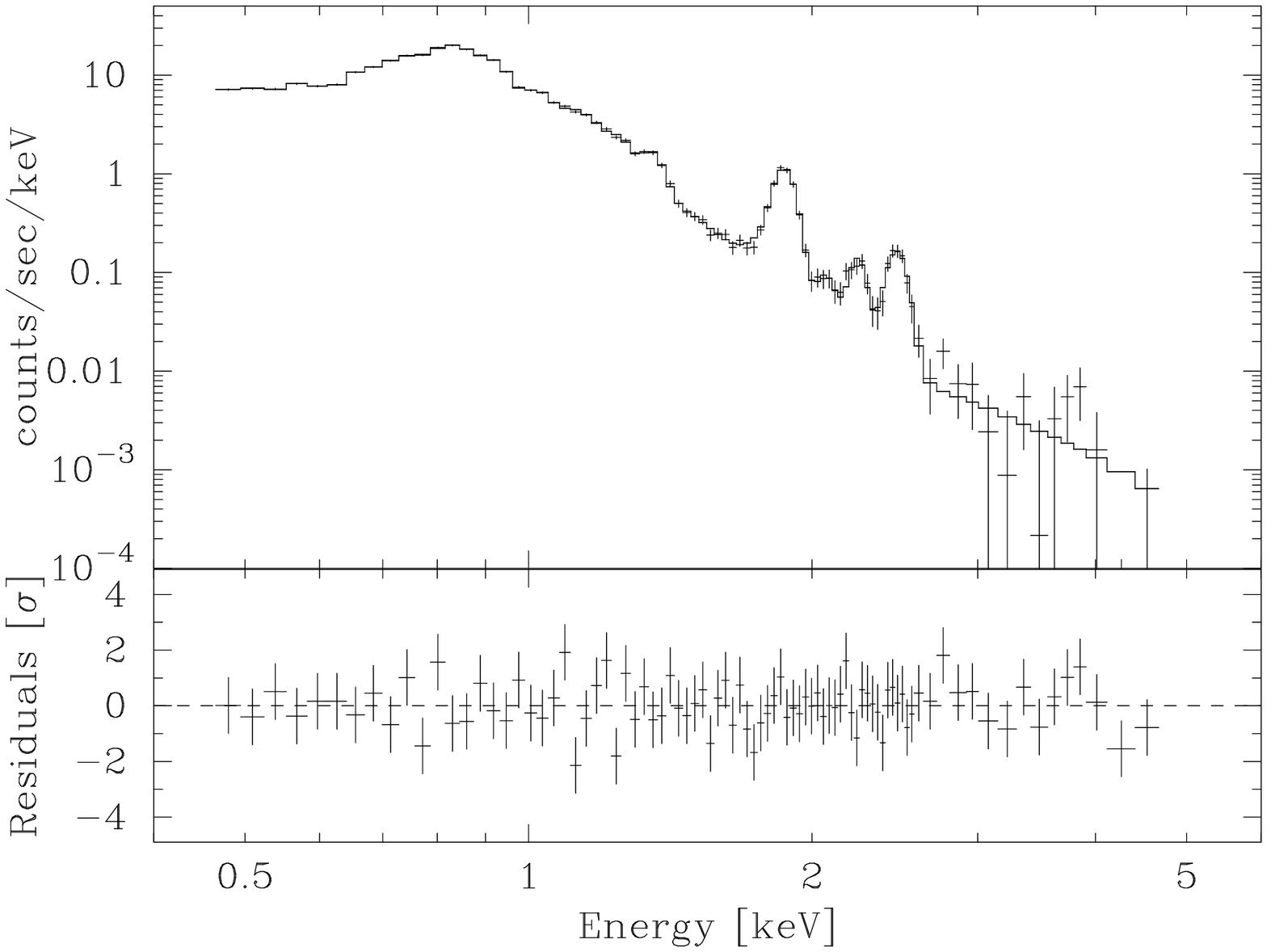}
\vspace{2cm}
\caption{}
\end{figure}

\begin{figure}[hptb]
  \centering
    \psbox[xsize=0.7#1]{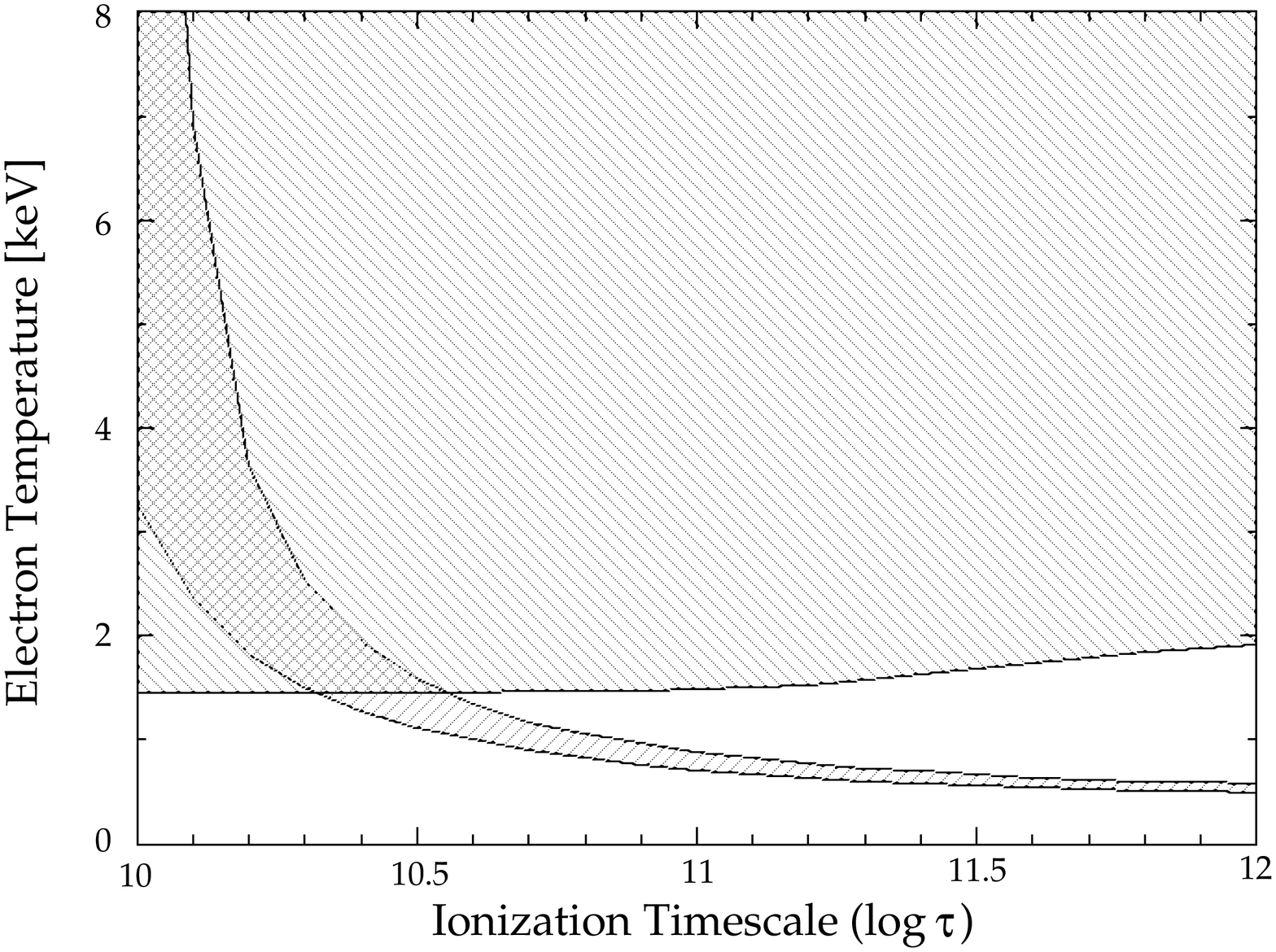}
\vspace{2cm}
\caption{}
\end{figure}

\begin{figure}[hptb]
  \centering
    \psbox{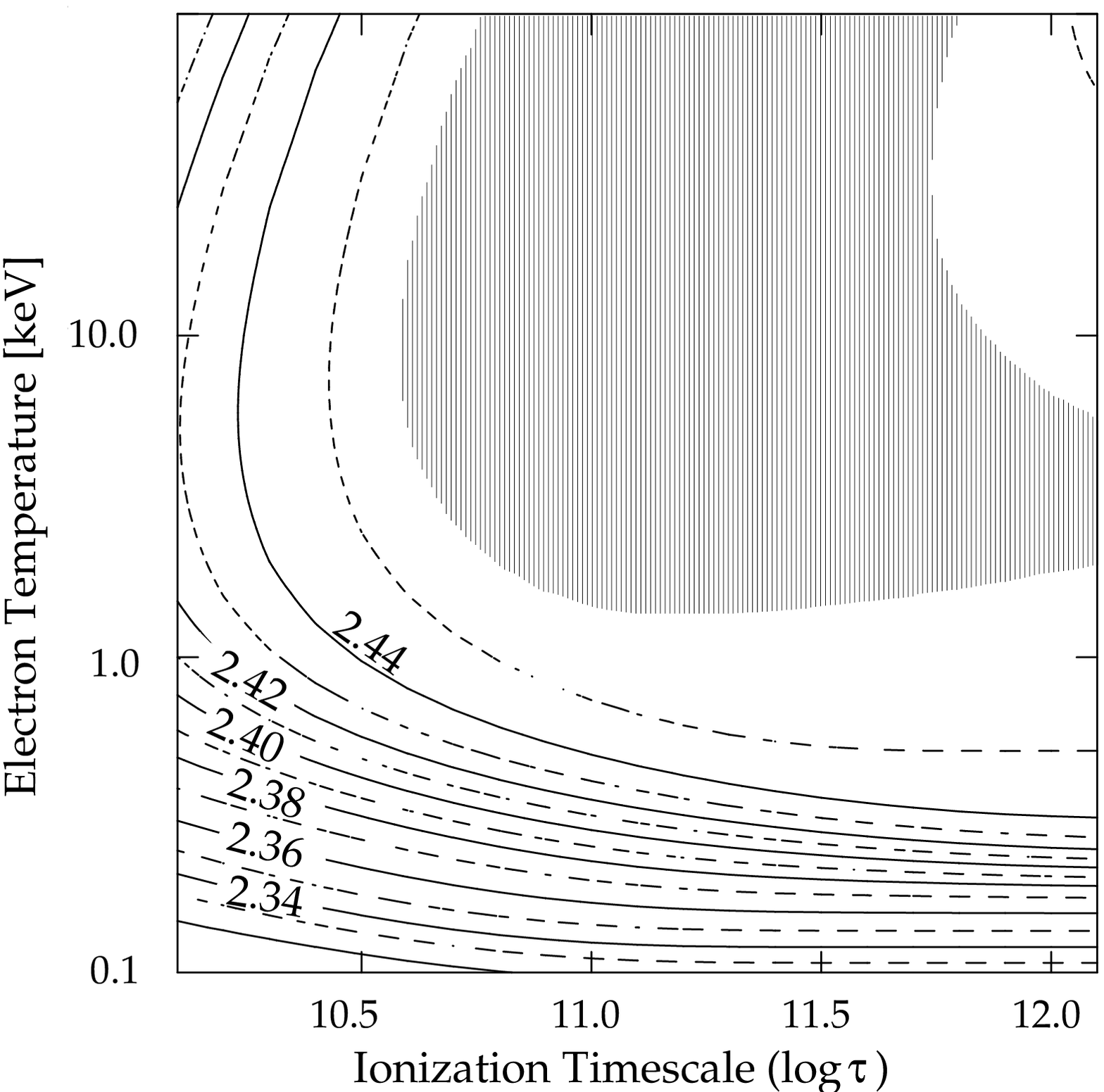}
\vspace{2cm}
\caption{}
\end{figure}

\begin{figure}[hptb]
  \centering
    \psbox[xsize=0.9#1]{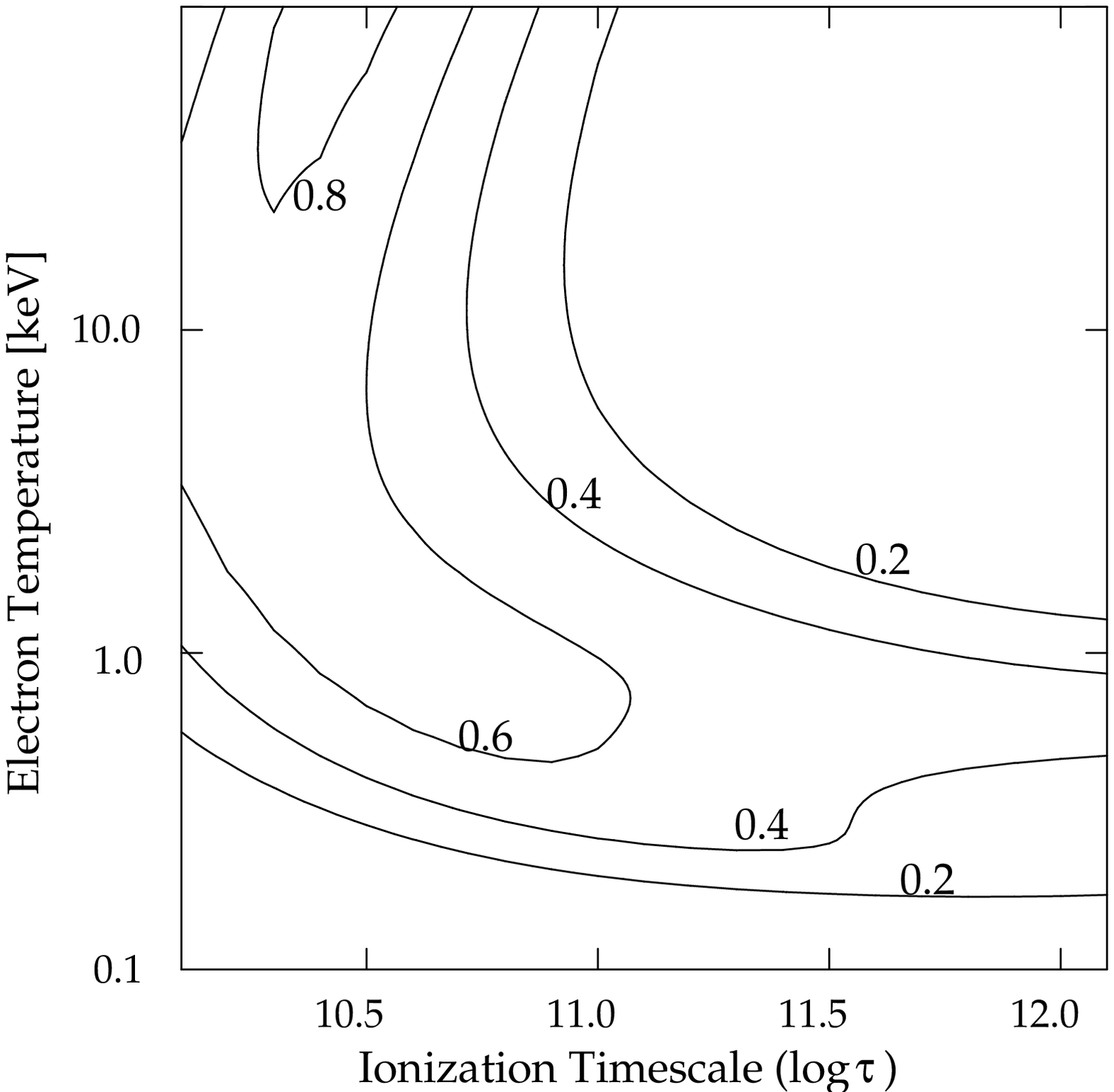}
\vspace{2cm}
\caption{}
\end{figure}

\begin{figure}[hptb]
  \centering
    \psbox[xsize=0.9#1]{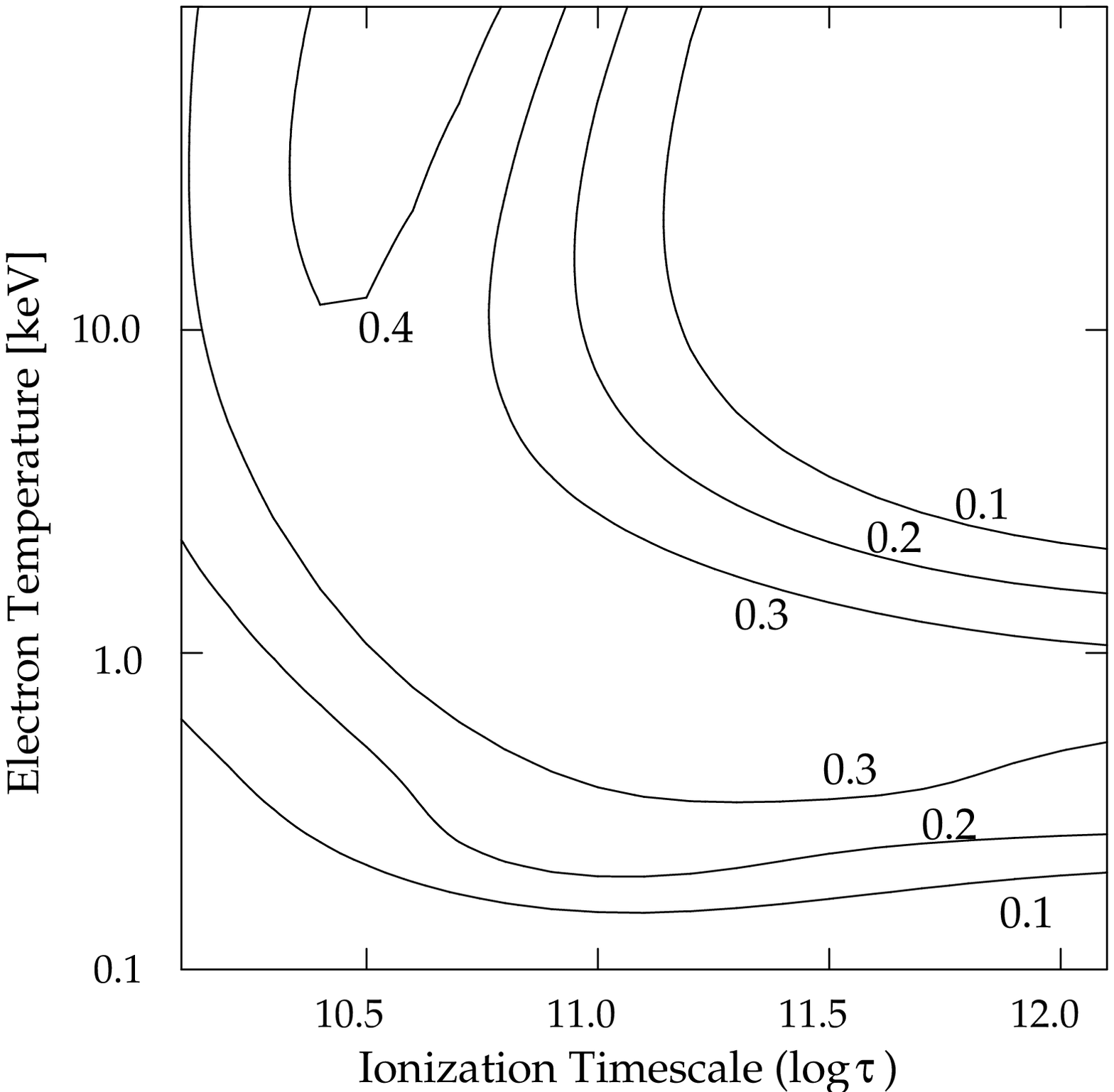}
\vspace{2cm}
\caption{}
\end{figure}

\begin{figure}[hptb]
  \centering
    \psbox[xsize=0.9#1]{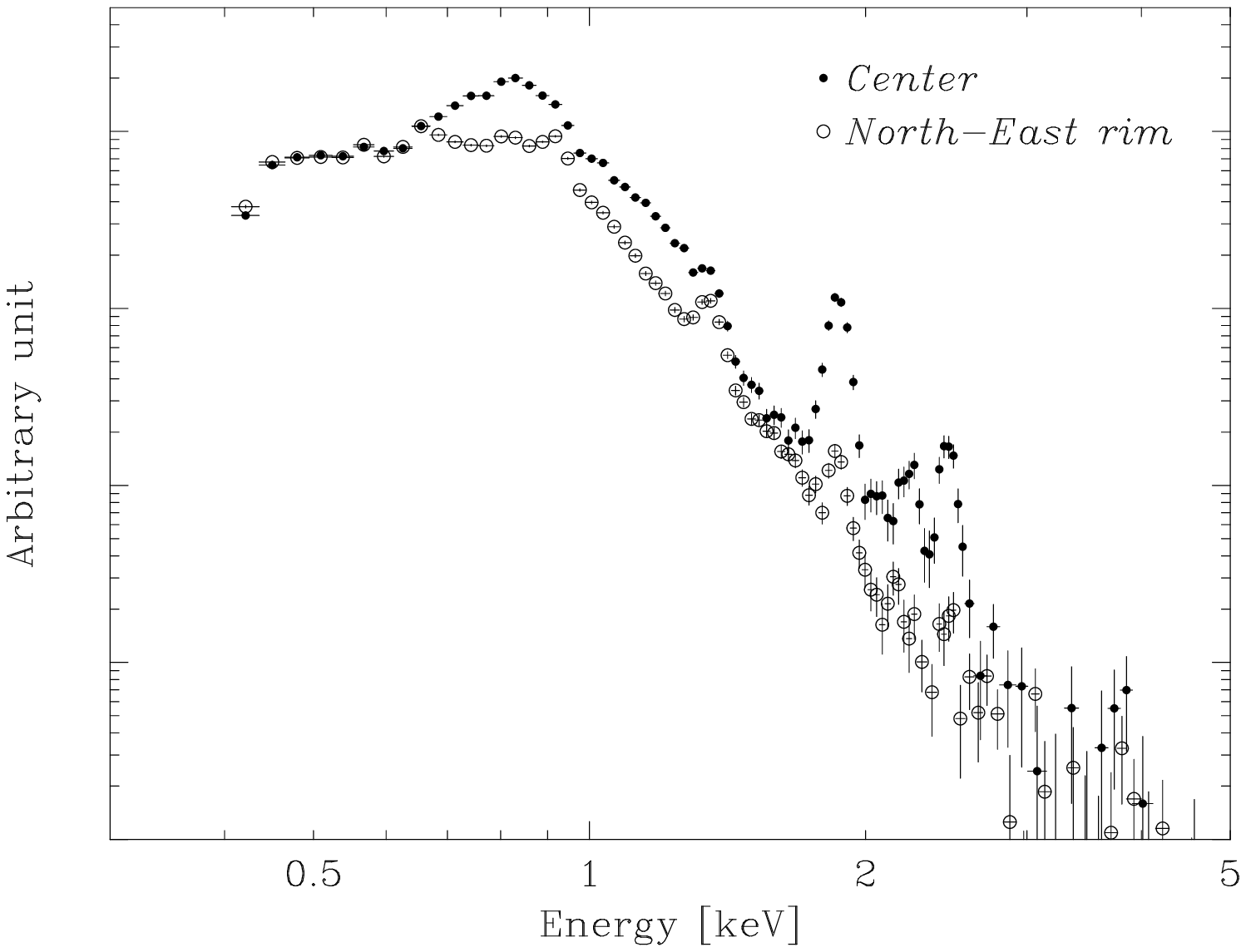}
\vspace{2cm}
\caption{}
\end{figure}

\begin{figure}[htbp]
    \centering
    \psbox[xsize=0.9#1]{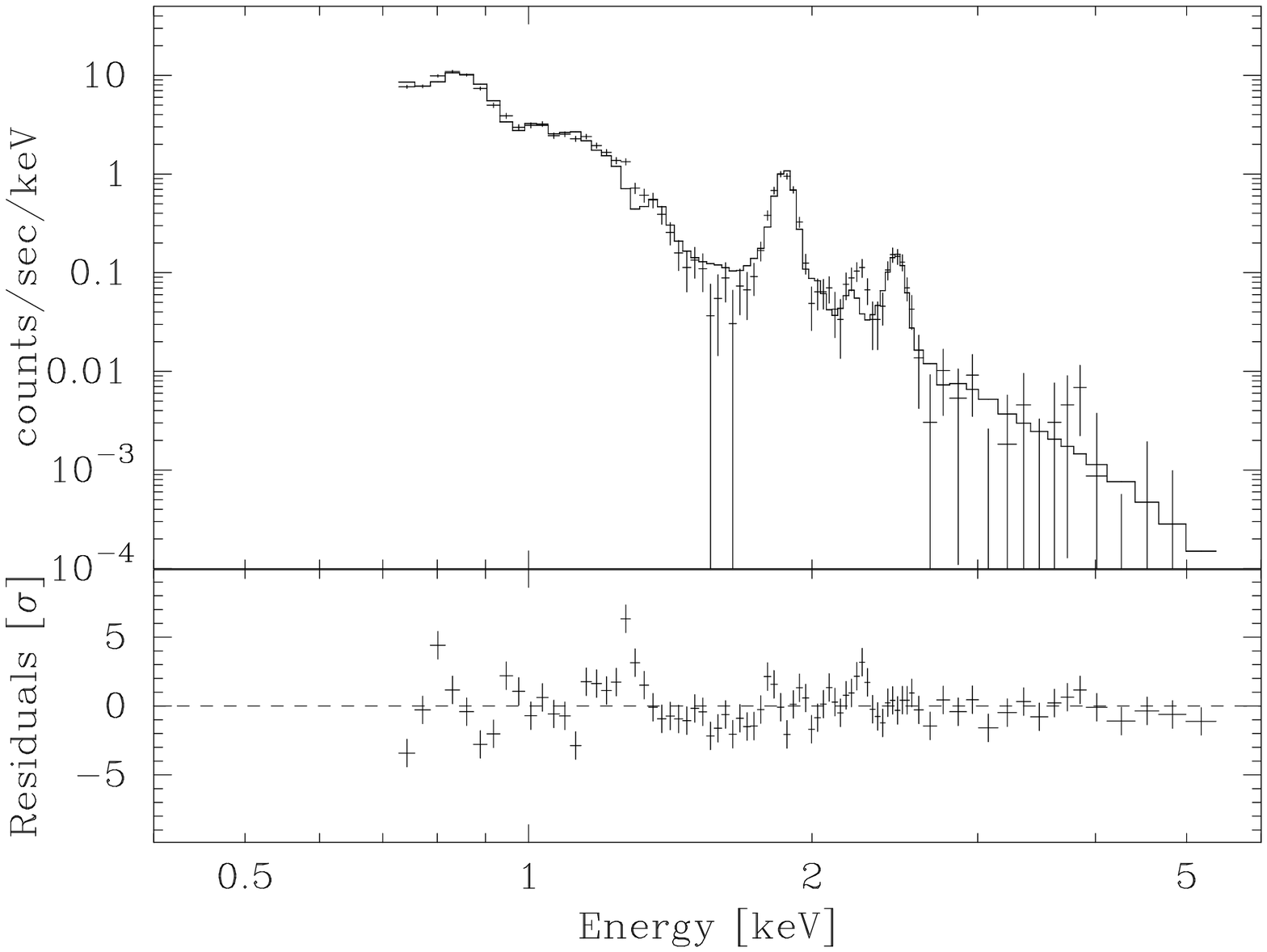}
\vspace{2cm}
\caption{}
\end{figure}

\begin{figure}[htbp]
    \centering
     \psbox[xsize=0.9#1]{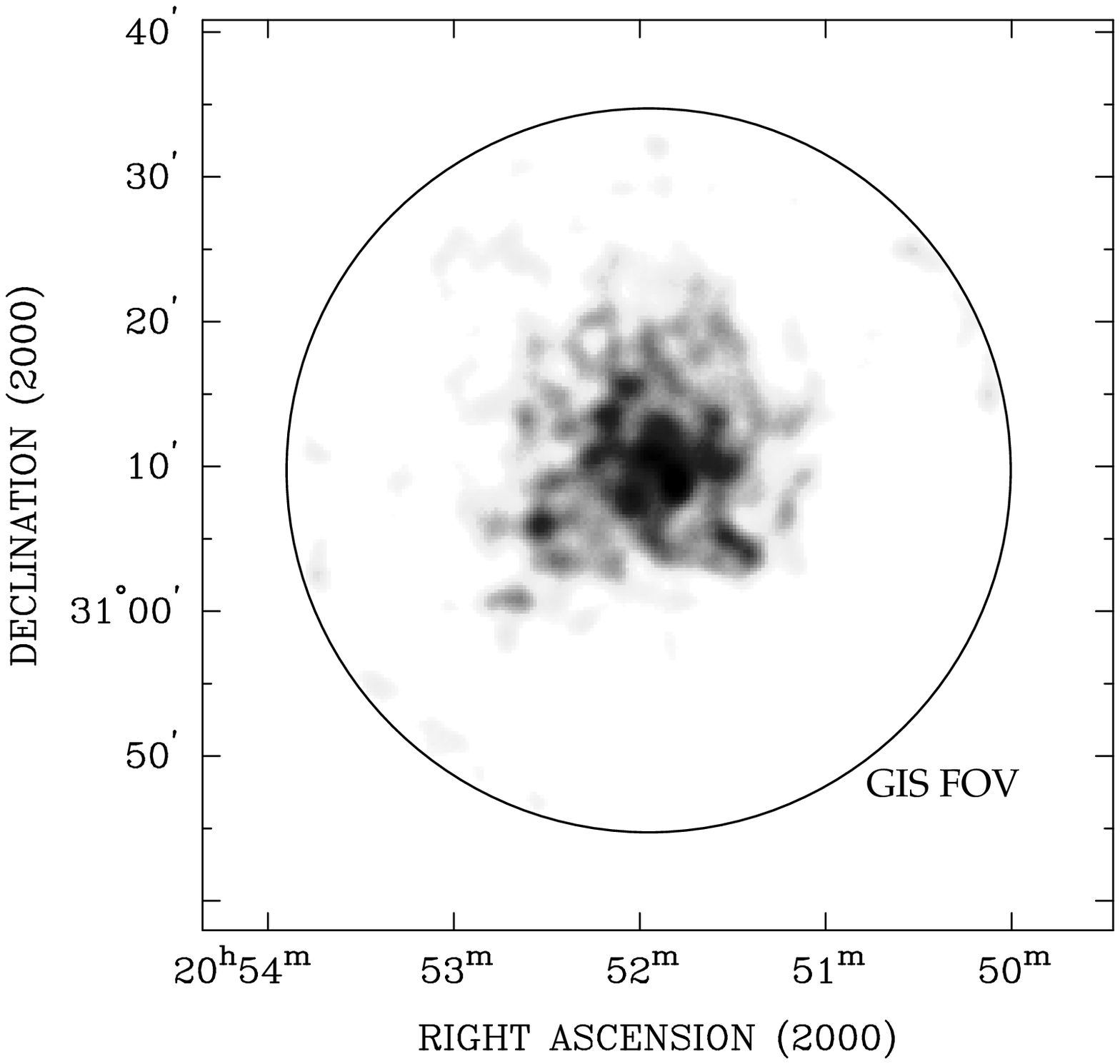}
\vspace{2cm}
\caption{}
\end{figure}

\begin{figure}[htbp]
    \centering
     \psbox[xsize=0.5#1]{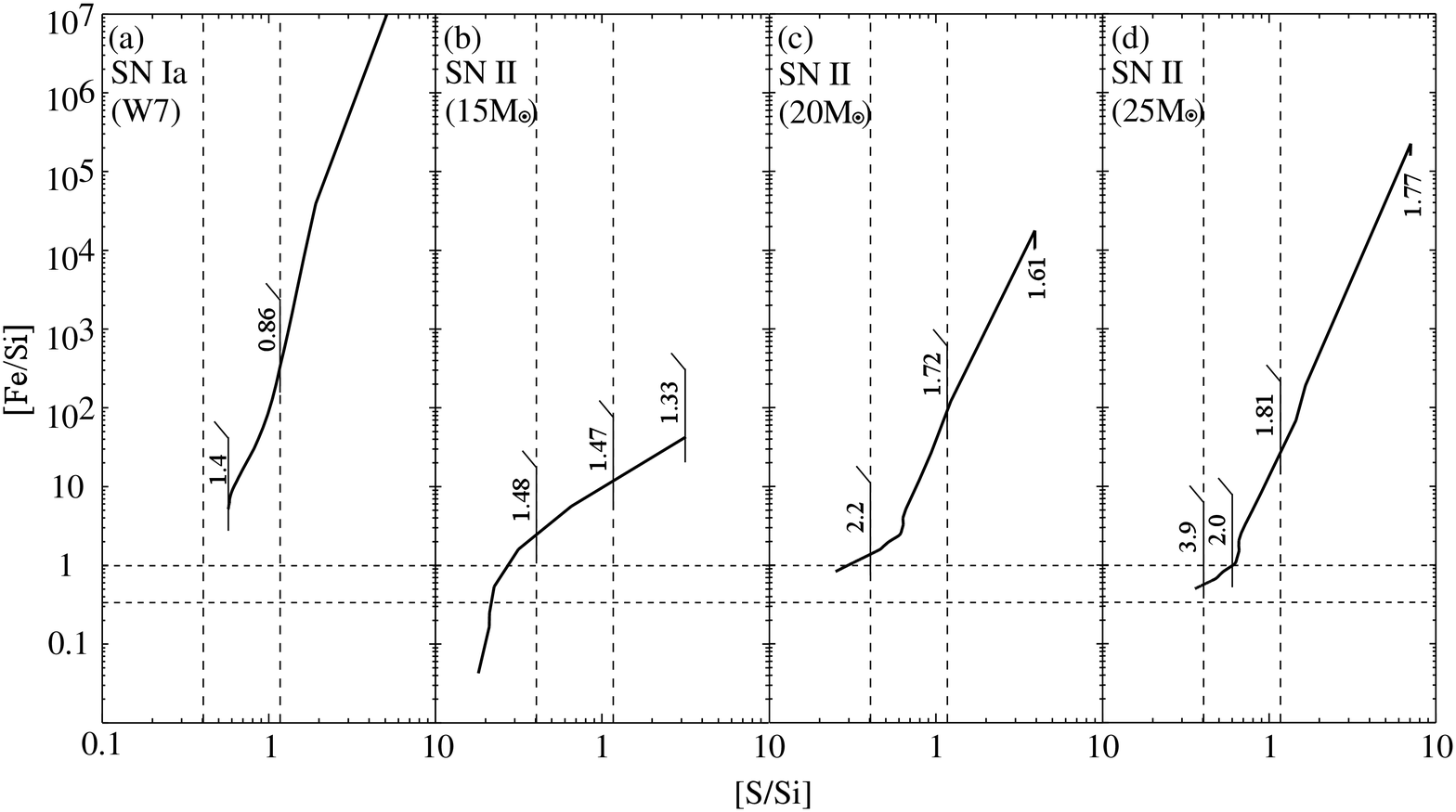}
\vspace{2cm}
\caption{}
\end{figure}

\begin{figure}[htbp]
  \centering
    \psbox{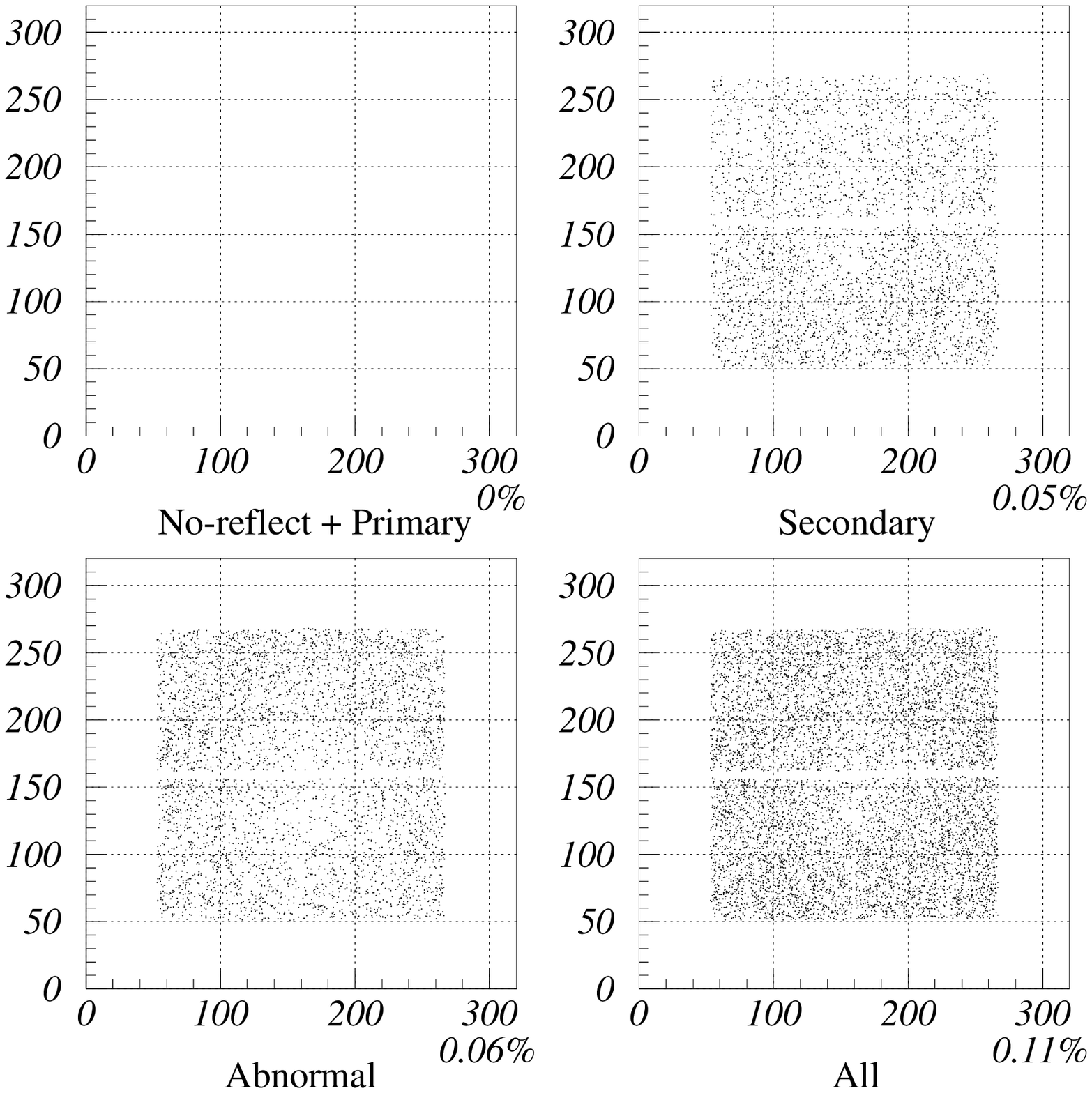}
\vspace{2cm}
\caption{}
\end{figure}

\end{document}